\documentclass[twoside,british,a4paper,twocolumn,bookmarks=true, colorlinks, linkcolor=blue, urlcolor=blue, citecolor=blue, p
lainpages=false, pdfpagelabels, final, breaklinks=true]{article}

\usepackage{amsmath}
\usepackage{amsfonts}
\usepackage{amssymb,upref}
\usepackage{tgheros} 
\usepackage{tgtermes} 
\usepackage{anyfontsize}
\usepackage{xcolor}
\usepackage{graphicx}
\usepackage{enumitem}
\usepackage{subfig}
\usepackage{siunitx}

\graphicspath{{./}{../}{../../}}

\usepackage{balance}

\usepackage[utf8]{inputenc}
\usepackage[T1]{fontenc}

\usepackage{isodate}
\cleanlookdateon

\usepackage{orcidlink}

\usepackage{lettrine}
\pdfmapfile{=montserrat.map} 

\usepackage{lipsum}

\usepackage{newtxmath}

\usepackage{geometry}
\usepackage{fancyhdr}

\fancypagestyle{firstpagestyle}
\usepackage{titlesec}
\titleformat{\section}[block]{\bfseries\upshape\sffamily\boldmath}{}{0.em}{}
\titleformat{\subsection}[block]{\upshape\sffamily\boldmath}{}{0.em}{}
\titlespacing*{\section}{0pt}{0.8em plus 0ex minus 0ex}{0em plus 0.ex}

\usepackage{titling}
\pretitle{\noindent\begin{minipage}{\textwidth} \raggedright\huge\bfseries\upshape\sffamily\boldmath
\color[rgb]{0,0,0.8}
}
\posttitle{\end{minipage} \vskip 0.6em}
\preauthor{\noindent \begin{minipage}{\textwidth} \large\mdseries\sffamily}
\postauthor{
  \end{minipage} 
  \vskip 0.4em
  {
   \sffamily
   \color{black!80}
   \noindent \small
   \address
   \par 
   \authoremail
   }
   \par \vskip 0.4em
  }
\predate{ \noindent \hspace{-0.4em} \sffamily\small}
\postdate{\vskip 0.0em}

\usepackage{caption}
\DeclareCaptionFont{cfs}{\fontsize{8.5}{10.25}\selectfont}
\DeclareCaptionLabelSeparator{vline}{\;|\;}
\usepackage{tcolorbox}
\definecolor{abstractboxcolor}{cmyk}{0.1,0,0,0}
\newtcolorbox{abstractbox}{
  arc=0pt,
  boxrule=0pt,
  colback=abstractboxcolor,
  boxsep=0.5em,
  left=0pt, right=0pt, bottom=0pt, top=0pt,
  width=\columnwidth
}
\makeatletter
 \def\@textbottom{\vskip \z@ \@plus 1pt}
 \let\@texttop\relax
\makeatother
\renewenvironment{abstract}{
   \noindent
   \begin{minipage}{\textwidth}
   \upshape\sffamily \bfseries
   \fontsize{9}{11.5}\selectfont
  }{
   \end{minipage} 
   \vskip 2.0em
  }

\usepackage[numbers,sort&compress]{natbib}

\makeatletter \def\NAT@def@citea{\def\@citea{\NAT@separator\,}} \makeatother 
\usepackage{etoolbox}
\apptocmd{\sloppy}{\hbadness 10000\relax}{}{}

\title{Chiral topological light for detection of robust enantiosensitive observables}
\newcommand\shorttitle{Chiral topological light for detection of robust enantiosensitive observables}

\author{
  Nicola Mayer\textsuperscript{1,2$\,*$}\orcidlink{https://orcid.org/0000-0002-8547-0073},
  David Ayuso\textsuperscript{1,3}\orcidlink{https://orcid.org/0000-0002-5394-5361},
  Piero Decleva\textsuperscript{4}\orcidlink{https://orcid.org/0000-0002-7322-887X},
  Margarita Khokhlova\textsuperscript{1,2}\orcidlink{https://orcid.org/0000-0002-5687-487X},
  Emilio Pisanty\textsuperscript{2}\orcidlink{https://orcid.org/0000-0003-0598-8524},
  Misha Ivanov\textsuperscript{1,3,5,6}\orcidlink{https://orcid.org/0000-0002-8817-2469}
  and
  Olga Smirnova\textsuperscript{1,6,7}\orcidlink{https://orcid.org/0000-0002-7746-5733}
  }
\newcommand\shortauthor{N.\ Mayer et al.}

\newcommand\address{
\textsuperscript{1} Max-Born-Institut, Max-Born-Str. 2A, Berlin, 12489 Germany \\
\textsuperscript{2} Attosecond Quantum Physics Laboratory, Physics Department, King's College London, Strand, London WC2R 2LS, UK \\
\textsuperscript{3} Department of Physics, Imperial College London, Prince Consort Rd, London, SW7 2NW, United Kingdom \\
\textsuperscript{4} CNR IOM and Dipartimento di Scienze Chimiche e Farmaceutiche, Universit\'a degli Studi di Trieste, Via Licio Giorgieri 1, Trieste, 34127, Italy \\
\textsuperscript{5} Department of Physics, Humboldt Universit\"at zu Berlin, Newtonstr. 15, Berlin, D-12489, Germany \\
\textsuperscript{6} Technion~-- Israel Insitute of Technology, 3200003, Haifa, Israel \\
\textsuperscript{7} Technische Universit\"at Berlin, Straße des 17.\ Juni 135, Berlin, 10623, Germany
}
\newcommand\authoremail{$^*$nicola.1.mayer@kcl.ac.uk}

\date{ \today }

\usepackage{textcomp}

\begin{document}

\twocolumn[
\begin{@twocolumnfalse}

\maketitle
\thispagestyle{firstpagestyle}

\begin{abstract}
The topological response of matter to electromagnetic fields is a property in high demand in materials design and metrology due to its robustness against noise and decoherence, stimulating recent advances in ultrafast photonics. 
Embedding topological properties into the enantiosensitive optical response of chiral molecules could therefore enhance the efficiency and robustness of chiral optical discrimination.
Here we achieve such a topological embedding by introducing the concept of chiral topological light~-- a light beam which displays chirality locally, with an azimuthal distribution of its handedness described globally by a topological charge.
The topological charge is mapped onto the azimuthal intensity modulation of the non-linear optical response, where enantiosensitivity is encoded into its spatial rotation.
The spatial rotation is robust against intensity fluctuations and imperfect local polarization states of the driving field.
Our theoretical results show that chiral topological light enables detection of percentage-level enantiomeric excesses in randomly oriented mixtures of chiral molecules, opening a way to new, extremely sensitive and robust chiro-optical spectroscopies with attosecond time resolution.
{
\vskip 0.3em
\normalfont \sffamily 
\footnotesize 
Accepted Author Manuscript.
Published online in \href{https://doi.org/10.1038/s41566-024-01499-8}{Nature Photonics (2024)} (in press),
\href{https://arxiv.org/abs/2303.10932}{arXiv:2303.10932}.
Available under \href{https://creativecommons.org/licenses/by-sa/4.0/}{CC BY}.
}
\end{abstract}

\vspace{-2mm}

\end{@twocolumnfalse}
]

\lettrine[lines=3, lhang=0.15]{T}{\:} 
he topological properties of the electronic response to electromagnetic fields in solid state systems, as well as in photonic structures, are being actively harvested to obtain robust observables, such as e.g.\ edge currents protected from material imperfections in topological insulators~\cite{RevModPhys.82.3045} or topologically protected light propagation pathways in their photonic analogs~\cite{Khanikaev:2013aa, rechtsman2013photonic}.
A similar robustness in the enantiosensitive optical response of gases or liquids of chiral molecules is strongly desired for analytical purposes, but is currently missing. While the first ideas connecting topological and chiral properties of electronic responses~\cite{ordonez2023geometric} or microwave signals in molecular gases~\cite{Kai:topological, Kai:microwave} are starting to emerge, they do not map onto the optical response, which encodes the ultrafast chiral electronic dynamics~\cite{Cireasa:2015aa}.

Topologically non-trivial optical signals can be achieved by using vortex beams, which carry orbital angular momentum~(OAM). They are characterized by an integer topological charge representing the number of helical revolutions of light's wavefront in space within one wavelength~\cite{Shen:2019aa}. Recent work established the chirality of vortex light in the linear regime~\cite{Forbes:2018, Forbes:2021}, exploited and manipulated ultrafast non-linear optical responses to vortex beams in atoms~\cite{Rego:2019ab, Dorney:2019aa}, including the discovery of new synthetic topologies~\cite{Pisanty:2019aa}, as well as in chiral molecules~\cite{Ashish:2023aa, Begin:2023aa}, where vortex light has also been successfully used for chiral detection in the hard X-ray region~\cite{Rouxel:2022aa, Rouxel:2022ab}. 
However, its natural enantiosensitivity in the optical domain is weak: the spatial scale of optical vortex beams is many orders of magnitude larger than the size of a molecule, making it difficult for the molecule to sense global
field structures.

This limitation can be overcome by encoding chirality 
of the optical field in time rather than in space. 
This means that locally, at a fixed point in space, the electric field vector of the electromagnetic wave draws a chiral three-dimensional Lissajous figure during one laser cycle.
Fields with such chiral Lissajous figures, referred to as "synthetic chiral light" \cite{Ayuso:2019aa}, employ only electric-dipole transitions to drive non-linear enantiosensitive signals. They have been devised~\cite{Kral:2001aa}, applied in the microwave region~\cite{Eibenberger:2017aa}, and extended to the optical domain~\cite{Ayuso:2019aa}.
The handedness of this light can be controlled with the phase delay between its frequency components, both locally --at every point in space -- and globally in the interaction region \cite{Ayuso:2022aa}.

Here we introduce the concept of chiral topological light, which takes advantage of the global topological structure of vortex light and the high enantiosensitivity of synthetic chiral light~\cite{Ayuso:2019aa}, embedding robust topological properties into the highly enantiosensitive ultrafast optical response.

\begin{figure*}[t]
\centering
\includegraphics[width=13cm, keepaspectratio=true]{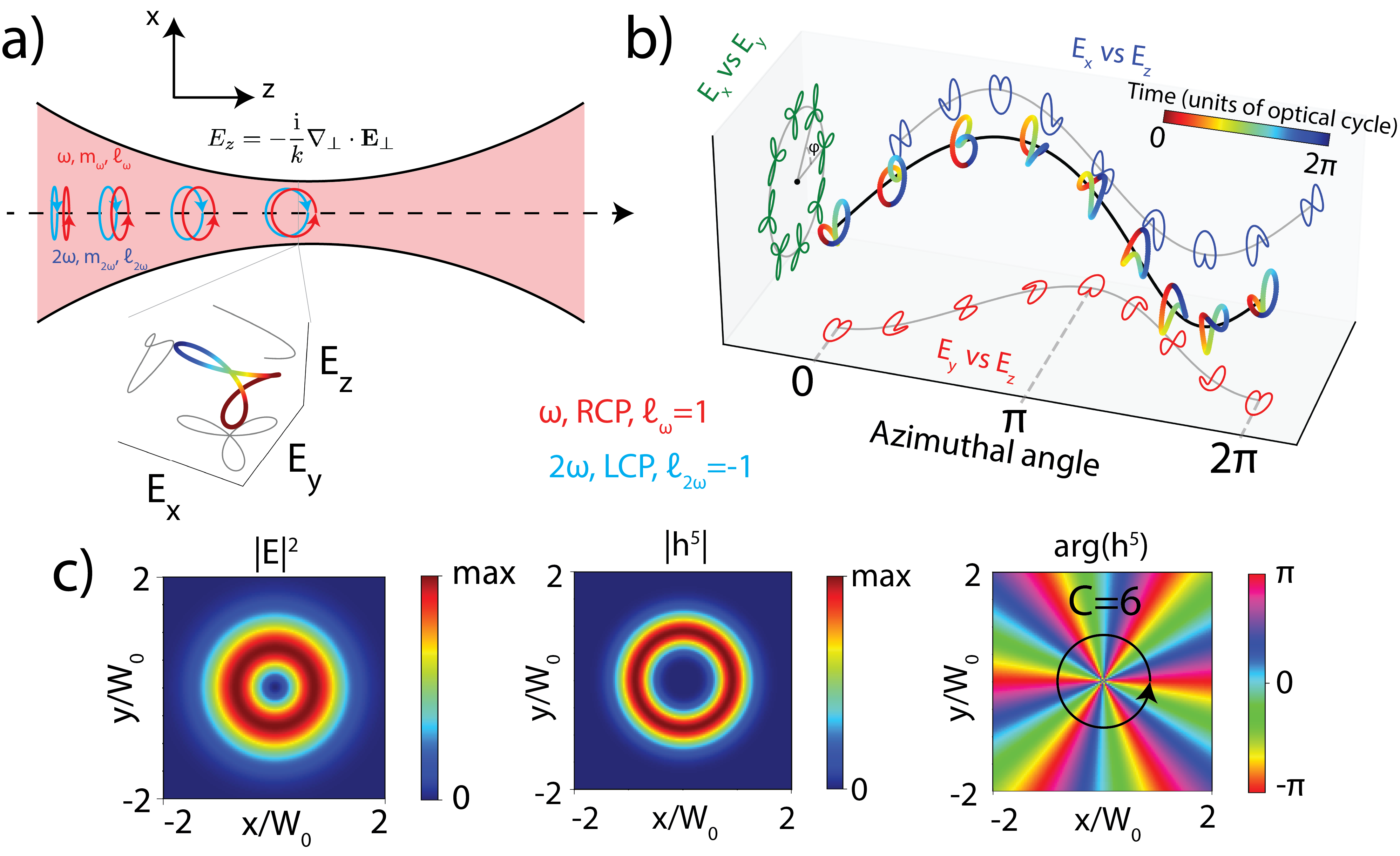}
\caption{The concept of chiral vortex light for bicircular counter-rotating $\sigma_{\omega}=-\sigma_{2\omega}=1$ beams carrying OAMs $\ell_{\omega}=-\ell_{2\omega}=1$. 
\textbf{a)}~Tight focusing of bicircular counter-rotating Gaussian beams induces a longitudinal field, resulting in a synthetic chiral field whose polarization vector draws a chiral Lissajous curve over one laser cycle (inset). 
\textbf{b)}~Evolution of the chiral Lissajous curves with respect to the azimuthal angle $\theta$ at a given radial position $\rho=\sqrt{x^2+y^2}$ at $z=0$ for a chiral vortex with $\ell_\omega=-\ell_{2\omega}=1$. \textbf{c)}~Slices through the electric field distribution at $z=0$. The figures show the total intensity of the electric field~$|\mathbf{E}|^2$, the absolute value of the chiral correlation function $|h^{(5)}|$ and its phase distribution $\arg\mathopen{}\left[h^{(5)}\right]\mathclose{}$. The phase distribution of $h^{(5)}$ describes the spatial distribution of the handedness of light and is characterized by a topological charge $C=6$. The $x$ and $y$ coordinates are scaled to the waist $W_0$ of the beams at the focus.}
\label{Fig1}
\end{figure*}

Our key idea is to imprint the topological properties of the vortex beam on the synthetic chiral light. Locally, the handedness of this light is characterized by the chiral correlation function $h$~\cite{Ayuso:2019aa}.
Thus, we aim to imprint the topological charge of the vortex beam on the azimuthal phase of $h$: $\arg[{h(\theta)}]=C\theta+\phi_L$. Here~$\theta$ is azimuthal angle, $C$ is the topological charge, and $\phi_L$ is the local enantiosensitive phase of the complex-valued correlation 
function $h$.

We now show that the intensity of the nonlinear optical emission of a chiral molecular medium triggered by such light depends on both chiral and topological phases of $h$ as well as the enantiosensitive phase $\phi_M$ introduced by the molecular medium: $I(\theta)\propto \cos(\phi_M-\phi_L+C\theta)$. 
We find that the azimuthal intensity profile is patterned in a topologically robust and molecule-specific way, leading to a large enantiosensitive offset $\Delta\theta=\pi/C$ between the intensity maxima (or minima) in opposite enantiomers. What's more, we find that the topologically controlled angular offset is robust with respect to imperfections of light polarization and intensity fluctuations, and  persists for very small amounts of enantiomeric excess.
Thus, it can be used to probe chirality in dilute mixtures.

To demonstrate these ideas, we focus on a specific realization of chiral topological light. It involves two Laguerre-Gaussian beams with counter-rotating circular polarizations, propagating along the $z$-axis with frequencies $\omega$ and $2\omega$ and OAMs $\ell_\omega$ and $\ell_{2\omega}$ (see Methods). Near the focus the field develops a longitudinal component given by $E_z=-(\text{i}/k)\nabla_{\perp}\cdot\mathbf{E}_{\perp}$ in the first post-paraxial approximation~\cite{Bliokh:2015aa} (see Fig.~\ref{Fig1}a), taking the light polarization vector out of the $(x,y)$ plane~-- a prerequisite for creating synthetic chiral light.

As a result, the Lissajous figure drawn by the polarization vector in one point in space over a laser cycle becomes chiral (see inset in Fig.~\ref{Fig1}a). Its handedness is controlled by the two-color phase $\phi_{2\omega,\omega}=2\phi_\omega-\phi_{2\omega}$, which depends on the azimuthal coordinate, forming a chiral vortex with the topological charge (see Methods): 
\begin{equation}\label{Eq1}C=2\ell_{\omega}-\ell_{2\omega}+2\sigma_{\omega}-\sigma_{2\omega}.\end{equation}
Here $\sigma_{r\omega}$ indicates right ($\sigma_{r\omega}=1$) or left ($\sigma_{r\omega}=-1$) circular polarization. The Lissajous curve drawn by the polarization vector of the electric field over one laser cycle changes with the azimuthal angle, switching handedness $2|C|$ times as the azimuthal angle cycles over one revolution (Fig.~\ref{Fig1}b). Thus, the superposition of two tightly-focused OAM beams at commensurate frequencies gives rise to a chiral vortex, i.e.\ a vortex beam displaying chirality locally at each given point with an azimuthally varying handedness characterized by an integer topological charge $C$.

Figure~\ref{Fig1}c visualizes the chiral vortex by displaying the beam total intensity $|\mathbf{E}(x,y)|^2$, the absolute value $|h^{(5)}(x,y)|$ and the phase $\arg[h^{(5)}(x,y)]$ of the chiral correlation function for OAM $(\ell_\omega,\ell_{2\omega})=\left(1,-1\right)$ and SAM $(\sigma_{\omega},\sigma_{2\omega})=(1,-1)$. Both the chirality and the total intensity maximize along rings (see Fig.~\ref{Fig1}c), typical for vortex beams, while the topological charge $C=6$ characterizes the azimuthal phase distribution of the light's handedness quantified by the chiral correlation function.

The chiral topological charge $C$ is highly tunable thanks to its dependence on the OAM of the two beams, which can take any integer value from $-\infty$ to $\infty$, enabling chiral vortices with arbitrarily high (and also arbitrarily low) chiral topological charge. 
By controlling the OAM of the beams, we can thus create chiral vortex beams with controlled properties. If $C=0$, then the chiral vortex has the same local handedness everywhere in space. Otherwise, the field's handedness displays a non-trivial spatial structure which is characterized by $C$.

\begin{figure*}[t]
\centering
\includegraphics[width=12cm, keepaspectratio=true]{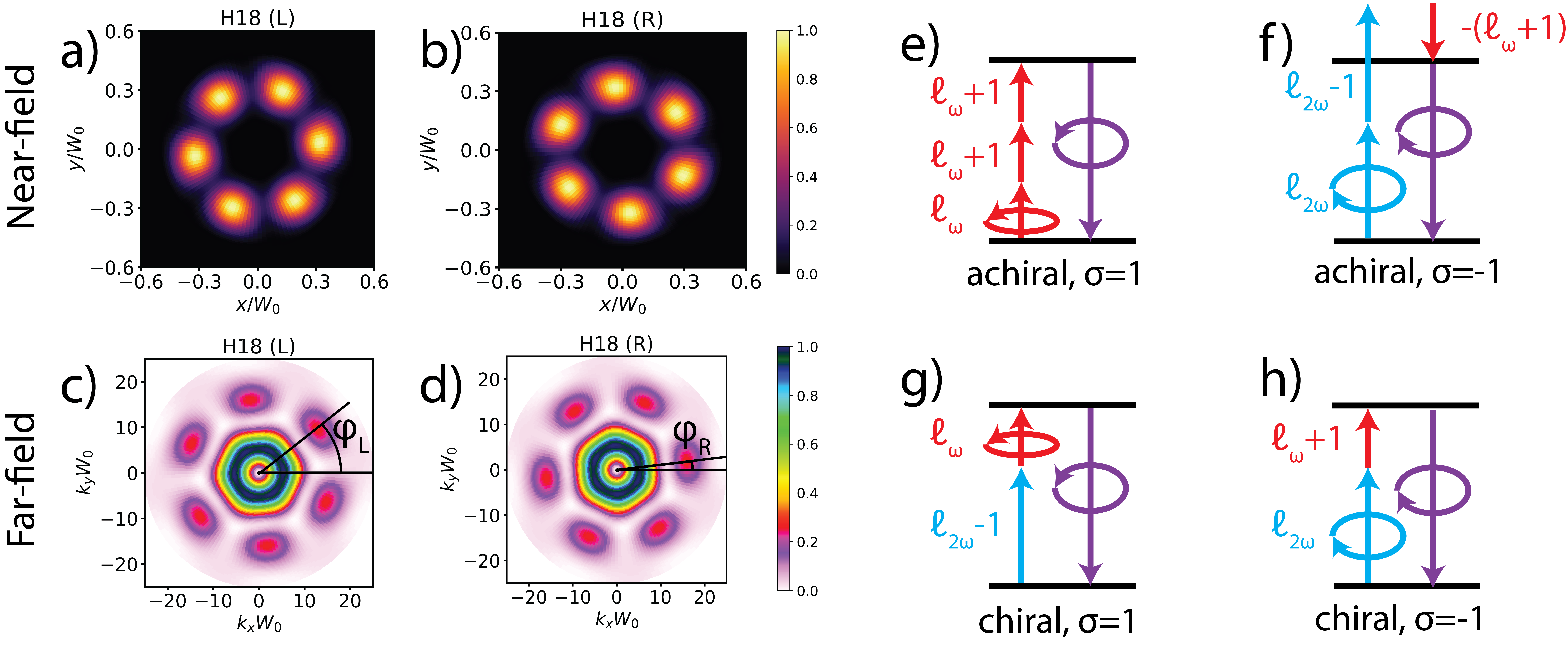}
\caption{Enantiosensitive high-harmonic spectroscopy using chiral topological light with topological charge $C=6$.
\textbf{a,b)}~The near-field spatial profile of H18 for L-fenchone~(\textbf{a}) and R-fenchone~(\textbf{b}). The $x$ and $y$ axes are given in units of the field waist at the focus $W_0$. 
\textbf{c,d)}~The corresponding far-field spatial profiles for the two enantiomers are shown in \textbf{c} for L-Fenchone and \textbf{d} for R-Fenchone. Here $k_x$ and $k_y$ are given in units of the reciprocal waist of the field at the focus, $1/W_0$. All profiles are normalized to their maximum value, which is the same for opposite enantiomers. The angles in the far-field picture indicate the position of the first peak in the outer ring of the profile, where we set the zero angle along the positive $k_x$ direction. For $C=6$ we have that $\phi_L=\phi_R+\pi/3$. \textbf{e-h)} Multiphoton diagrams describing the generation of $3N$ high-harmonic orders in a chiral molecule driven by chiral topological light. Photons carrying SAM $\sigma=\pm1$ are indicated with co- or counter-clockwise arrows, whereas longitudinally polarized photons are simple arrows. The $\ell_{r\omega}$ term corresponds to the OAM carried by each photon. 
\textbf{e,f)} The achiral channels (odd number of photons), where \textbf{e} corresponds to a final SAM of $\sigma=1$ and \textbf{f} to $\sigma=-1$. \textbf{g,h)} The chiral channels (even number of photons), for a final SAM of $\sigma=1$ (\textbf{g}) or $\sigma=-1$ (\textbf{h}). \textbf{e-g} A final SAM of the $3N$ order of $\sigma=1$ (\textbf{e,g}) or $\sigma=-1$ (\textbf{f,h}).}
\label{Fig2}
\end{figure*}

We have modeled the highly nonlinear response of randomly oriented chiral molecules to this realization of chiral topological light depicted in Fig.~\ref{Fig1} using a DFT-based S-matrix approach (see Methods).
Figure~\ref{Fig2}a,b shows the near-field intensity of harmonic 18 generated in R- and L-fenchone for fundamental frequency $\omega=0.044$~a.u.\ (1033 nm), peak intensity $I_0=5\cdot10^{14}$~W/cm$^2$ and a beam waist of $W_0=2.5$~$\mu\text{m}$ at the jet position $z=0$.

The azimuthal distribution of the near-field intensity records both the topology of the driving laser field and the handedness of the medium.
This azimuthal distribution results from the interference between chiral and achiral multiphoton pathways.
The maxima occur at angles $\theta=\left[2\pi n+(\phi_L-\phi_M)\right]/C$, where the two pathways interfere constructively. The angular position of the peaks is therefore enantiosensitive: swapping the molecular enantiomer leads to a $\pi$ shift in the molecular phase $\phi_M\rightarrow\phi_M+\pi$, shifting the minima and maxima of the intensity pattern by $\pi/C$. The number of peaks is controlled by the topological charge $|C|=6$.
Importantly, the same topological structure is preserved in the far-field response, see Fig.~\ref{Fig2}c,d.
In the multiphoton picture of HHG, the enantiosensitive topological structure arises due to the interference between achiral and chiral channels \cite{Ayuso:2019aa}, which we depict in Fig.~\ref{Fig2} e,f,g,h. Taking into account both the SAM $\sigma$ and OAM $\ell$ carried by each photon, it is easy to see that the difference in net OAM transferred to the harmonic orders in chiral and achiral channels corresponds to the topological charge $C$ (see Methods for a detailed description).
\begin{figure*}[t]
\centering
\includegraphics[width=13cm, keepaspectratio=true]{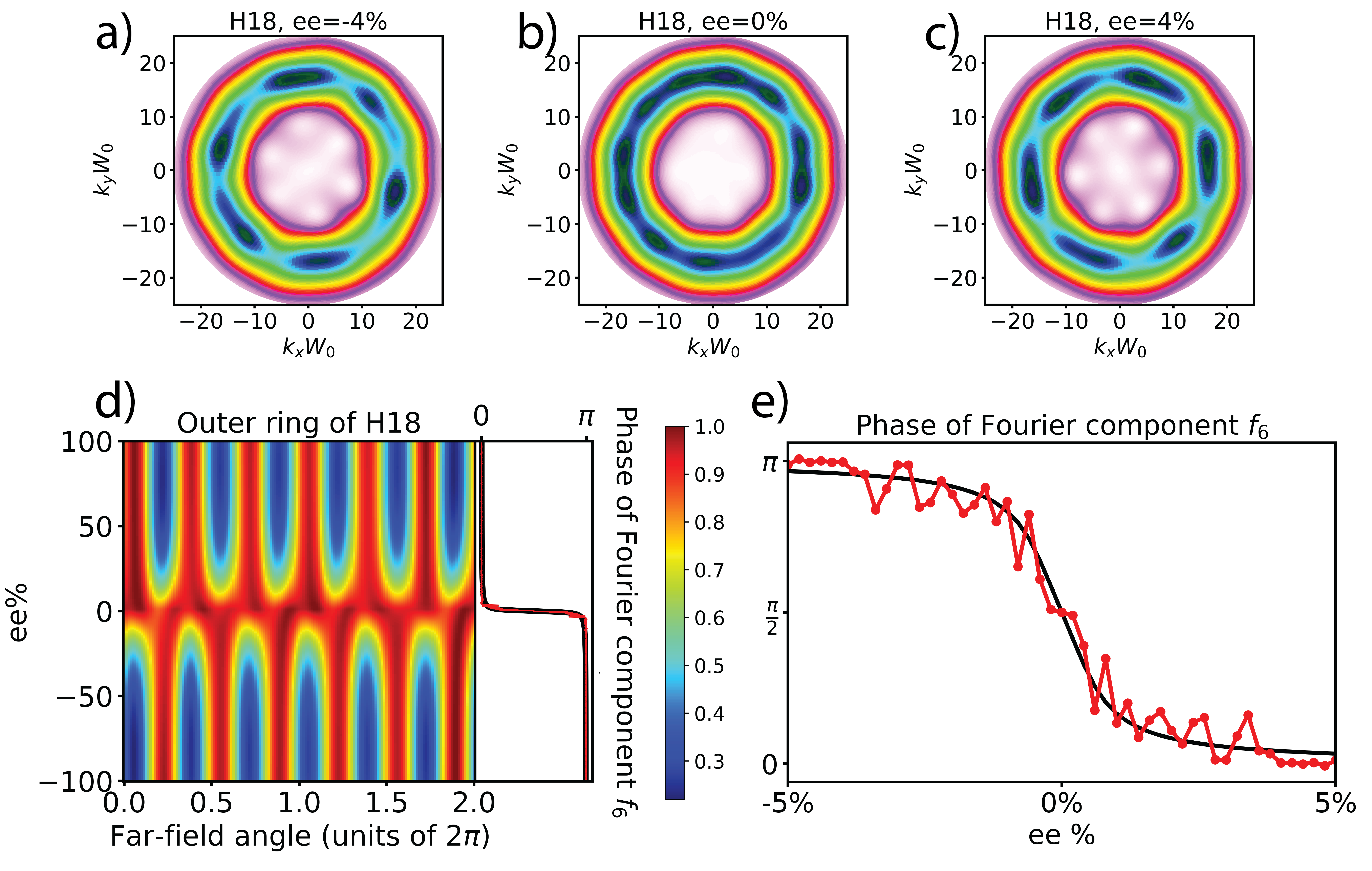}
\caption{Dependence on the enantiomeric excess. 
\textbf{a-c)}~The far-field spatial profiles of H18 for an enantiomeric excess $ee=(N_R-N_L)/(N_R+N_L)$ of -4\% (\textbf{a}), 0\% (\textbf{b}) and 4\% (\textbf{c}), where a positive e.e. corresponds to a larger concentration of R-fenchone in the sample. 
\textbf{d)}~Angle-resolved, radially-integrated far-field signal of the outer ring of the spatial profile ($|kW_0|>10$) as a function of e.e. The black line on the right shows the phase of the Fourier component of the spatial profile oscillating at frequency $\ell=6$ as a function of e.e. The overlapping red line shows the result accounting for intensity fluctuations. The $\pi$ jump at $ee=0\%$ indicates the enantiosensitive rotation of the spatial profile. 
\textbf{e)}~The phase of the Fourier component is shown with black solid line. The red line with circles shows the phase obtained including intensity fluctuations for e.e. between -5\% and 5\%.}
\label{Fig3}
\end{figure*}

Encoding the topological charge $C$ into the molecular response and extracting the enantiosensitive offset angle, controlled by $C$, allows us to measure the enantiomeric excess $ee=(N_R-N_L)/(N_R+N_L)$ in macroscopic mixtures of left- and right-handed molecules with concentrations $N_L$ and $N_{R}$. 
Even for very small values of $ee$, we observe the appearance of the $C$-fold structure in the inner and outer rings, as well as the corresponding enantiosensitive rotation of the spatial profile, see Figs.~\ref{Fig3}a,c. For $ee=0\%$ shown in Fig.~\ref{Fig3}b chiral channels are suppressed, and a topologically different $2C$-fold structure is observed as a result of the interference between the two strongest open achiral channels leading to  H18 and allowed by the selection rules, depicted 
in Fig.\ref{Fig2} e,f. (see Methods and Supplementary Information,~SI, for details).

The enantiosensitive rotation of the $C$-fold structure in the outer ring is apparent in the angle-resolved, radially-integrated signal (Fig.~\ref{Fig3}d). It manifests in the abrupt switching of the azimuthal angle which maximizes the signal, as one changes the enantiomeric excess from positive to negative. The enantiosensitive rotation can be easily separated by performing a Fourier analysis of the signal with respect to the azimuthal angle as a function of the enantiomeric excess. The solid black line in Fig.~\ref{Fig3}d shows the phase of the Fourier component $f_6$ oscillating at the $C=6$ frequency of the outer ring signal, as a function of the enantiomeric excess. A clear $\pi$ phase jump is observed at $ee=0\%$, indicating the switch in the handedness of the mixture. The sharpness of this jump (Fig.~\ref{Fig3}e) characterizes the accuracy of resolving left- and right-handed molecules in mixtures with vanishingly small enantiomeric excess. 

We now show that the enantiosensitive signal is robust with respect to imperfections in the laser beams.
First, we include noise in our simulations (see Methods) via 2$\%$ intensity fluctuations of the driving fields. The red line in Figs.~\ref{Fig3}d,e shows the phase of the Fourier component $f_6$ when intensity fluctuations are included. It is clear that the sharp $\pi$ shift of the phase is robust against noise, see Fig.~\ref{Fig3}e. It allows us to distinguish positive and negative enantiomeric excess with high fidelity, on the scale $\sim 0.1\%$, demonstrating  topological robustness of the enantio-sensitive signal. The topological structure is imprinted via azimuthal interference of chiral and achiral responses and survives as long as the two-color phase remains stable, which is routinely achieved in two-color experiments with extremely high accuracy (see e.g. ~\cite{Fleischer:2014aa}). 
We thus expect robust encoding of topological information in the molecular gas and robust read-out of the chiral topological signal.

Fig.\ref{Fig3} is the first key result of our work: our method is sensitive to very small values of enantiomeric excess in nearly equal mixtures of left- and right-handed molecules.
This sensitivity, at the level well 
below 1$\%$, rivals or even exceeds the golden 
standard achieved in photo-electron spectroscopy \cite{Comby:2023aa,kastner2016enantiomeric} using the new generation of chiral-sensitive methods  relying on the electric dipole interactions
\cite{powis2000photoelectron, bowering2001asymmetry, lux2012circular, janssen2014detecting,nahon2015valence}.

Typical experimental imperfections are  related to the imperfect circularity (SAM) and imperfect OAM contents of the light beams. We show below that
even though such imperfections affect the topological charge, 
the concept of enantiosensitive rotation of the non-linear response remains valid. 

\begin{figure}[t]
\centering
\includegraphics[width=\columnwidth, keepaspectratio=true]{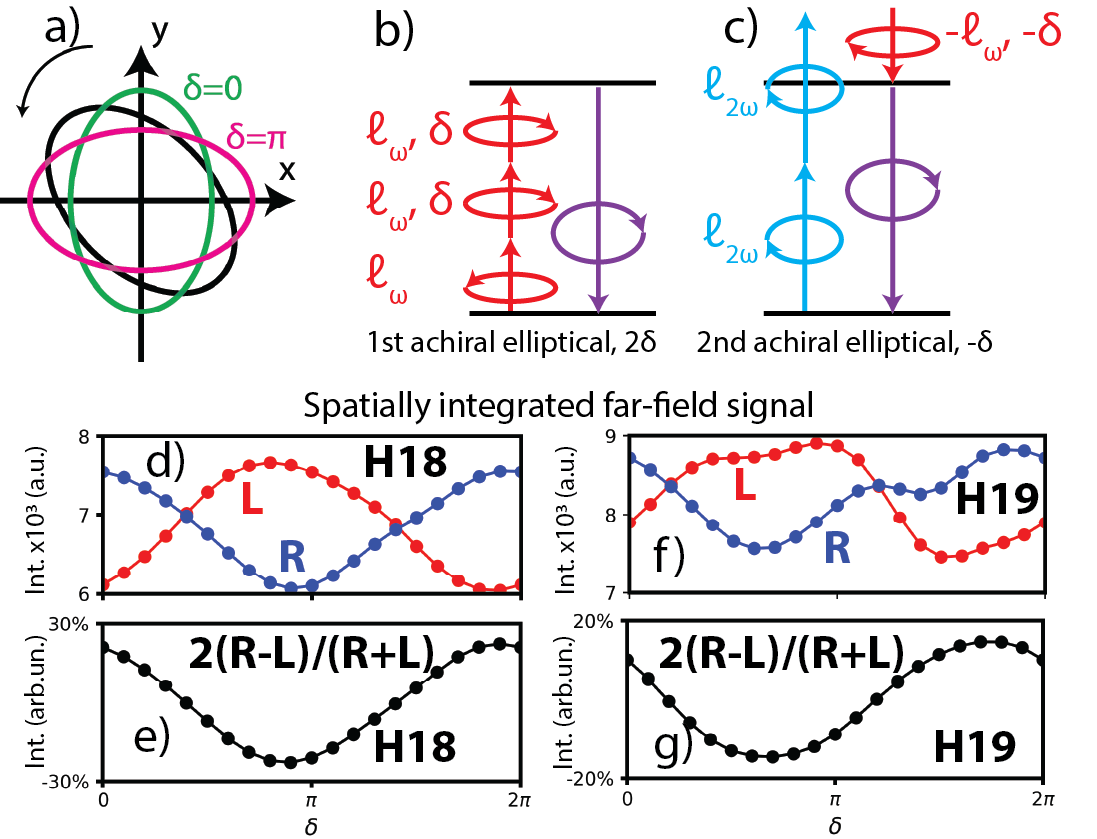}
\caption{Fourier analysis to recover the enantiosensitive rotation of the spatial profile in the case of elliptical fields. 
\textbf{a)}~Ellipse of an elliptical field and its orientation in the $(x,y)$ plane with respect to the phase delay $\delta$ between the counter-rotating components. \textbf{b,c)} Multiphoton diagrams of the new achiral channels contributing to the $3N$ harmonic orders for chiral topological light with elliptical $\omega$ field, where \textbf{b} corresponds to a final phase delay dependence of $2\delta$ and \textbf{c} to a final phase delay dependence of $-\delta$. The co- and counter-clockwise arrows indicate the SAM $\sigma$, whereas $\ell_\omega$ corresponds to the OAM; $\delta$ is the phase delay. \textbf{d-g)}~Spatially-integrated far-field signal for H18 (\textbf{d,e}) and H19 (\textbf{f,g}) as a function of the phase delay $\delta$. The red (blue) dotted lines in \textbf{d} and \textbf{f} corresponds to the signal $S_L$ ($S_R$) from L-(R-)fenchone, whereas the black dotted lines in \textbf{e} and \textbf{g} correspond to the chiral dichroism signal $2(S_R-S_L)/(S_R+S_L)$. Int., intensity. 
}
\label{Fig4}
\end{figure}

Consider chiral topological light created by elliptically polarized drivers with imperfect circularity (see SI for imperfections in the OAM content). 
To understand its effect on our observables, we express the elliptical field in terms of two counter-rotating circularly polarized components: $\mathbf{E}(\omega)=[(1+\epsilon)\mathbf{E}_+(\omega)$ $+$ $(1-\epsilon)\exp(\text{i}\delta)\mathbf{E}_-(\omega)]/\sqrt{2(1+\epsilon^2)}$. Here $\delta$ is the phase delay between the components, which corresponds to the orientation of the resulting elliptical polarization and can be well controlled in the experiment~\cite{Comby:2023aa}, and $|\epsilon|\le 1$ is the ellipticity, which is difficult to control with few-percent accuracy. Note that $\delta=0,\pi$ correspond to elliptical light ``squeezed'' along the $x$- and $y$-axis respectively (see Fig.~\ref{Fig4}a), and that both $\delta$ and $\epsilon$ can be related to standard Stokes parameters \cite{Stokes:2007}.

The appearance of the additional counter-rotating component in the elliptical beam leads to two interrelated consequences: (i)~the change of the topological structure of the harmonic emission due to the presence of new SAM components in the beams (see Eq.~\ref{Eq1}) leading to admixture of emission with topological charge C=-2, and (ii)~the appearance of two strong multiphoton pathways contributing to the achiral harmonic signal and effectively masking a weaker chiral signal driven by the longitudinal polarization. The new achiral multiphoton  channels arising due to 
imperfect circularity of the pulse are shown in Fig.~\ref{Fig4}b,c (see Methods for additional details). The contribution of different 
topological charges and different multiphoton pathways are disentangled by realizing an analogue of the lock-in method:
we use the dependence of the signal on $\delta$, i.e. 
on the orientation angle of the polarization ellipse. 
This dependence is different for the different 
nonlinear optical diagrams carrying different topological charges.
Hence, rotating the polarization ellipse, i.e. changing $\delta$, and Fourier transforming the signal with respect to $\delta$ allows us to decouple the different contributions to the  
signal  according to  their topological charges and consequently amplify the chiral signal.
\begin{figure*}[t]
\centering
\includegraphics[width=16cm, keepaspectratio=true]{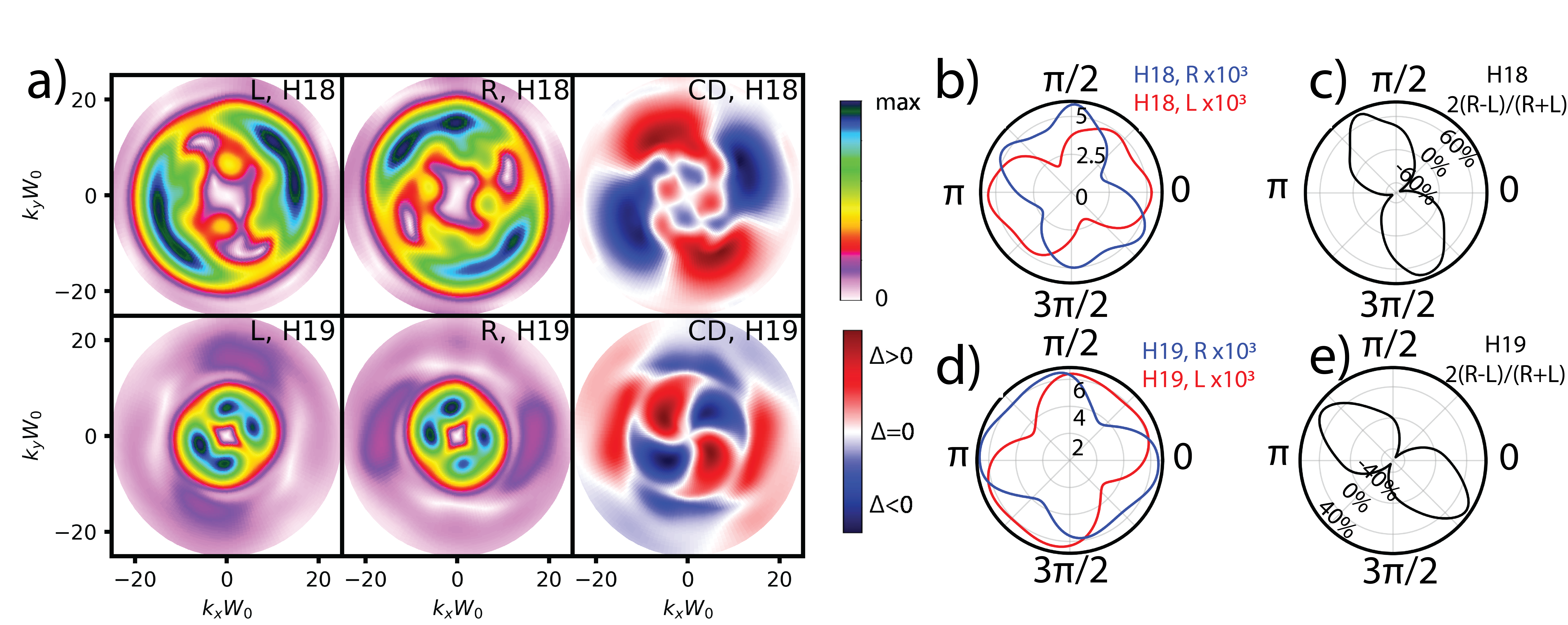}
\caption{\textbf{a)}~Far-field spatial profiles for H18 and H19 obtained after Fourier transform with respect to the phase delay $\delta$ at the $\tilde{\delta}=1$ Fourier component for H18 (top ) and H19 (bottom). The left (center) column shows the results for L-(R-)fenchone, whereas the right column shows the difference signal $S_R-S_L$. 
\textbf{b-e)}~Radially-integrated signal as a function of the azimuthal angle of the far-field spatial profiles for H18~(\textbf{b,c}) and H19~(\textbf{d,e}). The solid red (blue) lines in \textbf{b} \textbf{d} correspond to L-(R-)fenchone, whereas the black lines in \textbf{c} and \textbf{e} show the chiral dichroism signal $2(S_R-S_L)/(S_R+S_L)$.}
\label{Fig5}
\end{figure*}

The additional benefit of using elliptical drivers is the access to globally chiral light with non-zero topological charge leading to total, integrated over the spatial profile,  enantio-sensitive intensity in all harmonic orders, providing the opportunity to harvest not only the relatively weak $3N$ harmonics, but also the naturally intense $3N+1$ harmonics.  Figures~\ref{Fig4}d,e,f,g show the total far-field intensity for H18 (~\ref{Fig4}d,e) and H19 (~\ref{Fig4}f,g) as a function of the phase delay $\delta$ between the counter-rotating components of the $\omega$ field for both R- ($S_R$, blue dotted line) and L-fenchone ($S_L$, red dotted line), as well as the chiral dichroism~$2(S_R-S_L)/(S_R+S_L)$ (black dotted line), for globally chiral topological light with fundamental beam ellipticity of $\epsilon_\omega=0.9$.
The other parameters are kept as above. 
The strength of the far-field signal changes as one rotates the ellipse of the $\omega$ field, while the chiral dichroism in the signal intensity is maximized at around 20\% for harmonic 19 and around 30\% for harmonic 18.

Fourier-transforming the far-field intensity profile with respect to the phase delay $\delta$ separates the contributions of different pathways, because they experience different modulation with $\delta$. The two achiral pathways interfere in the third Fourier component ($\tilde\delta=3$, see Figs.~\ref{Fig4}b,c) with respect to $\delta$, while the dominant contribution between the chiral and achiral pathways corresponds to the first Fourier component ($\tilde\delta=1$). Figure~\ref{Fig5}a shows the far-field spatial profiles for $\tilde\delta=1$ for both enantiomers and both H18 and H19, as well as the their difference, while the polar plots in Fig.~\ref{Fig5}b,c,d,e show the radially-integrated signals and chiral dichroism. We see that the Fourier filtering recovers the enantiosensitive rotation, although the dominant topological charge is now $C=2$, (Eq.~\ref{Eq1} for $\sigma_\omega=\sigma_{2\omega}=-1$).

Fig.~\ref{Fig5} is the second key result of our work.  It  demonstrates a robust route to decomposing the contributions to the overall chiral optical signal, originating from interfering pathways encoding different topological charge. The decomposition relies on straightforward Fourier analysis of the far field image. Given the ability to precisely control the orientation of the polarization ellipse of the incident infrared light, chiral topological light generated by such infrared drivers stands out as a robust probe of molecular chirality, capable of inducing strongly enantiosensitive total intensity signals as well as giant rotations of intense spectral features.

The concept of chiral topological light introduced here is not limited to vortex beams: other members of the larger family of structured light beams~\cite{Angelsky:2020aa, Forbes:2021aa, Dunlop:2017aa} can be used to create locally and globally chiral topological light. We envision using tightly focused radially polarized beams, which are known to posses strong longitudinal components~\cite{Dorn:2003aa}, central to the concept of local chirality. Skyrmionic beams~\cite{Cuevas:2021aa, Du:2019aa} could also be used, e.g. to induce topological distributions with radially-dependent topological charges.
From the perspective of structured light~\cite{Angelsky:2020aa, Forbes:2021aa, Dunlop:2017aa, Bliokh:2023aa} the temporally chiral vortex introduced here represents a new kind of polarization singularity, which could be analyzed by extending the current framework from the monochromatic three dimensional fields~\cite{Bliokh:2019aa, Alonso:2023aa} to the polychromatic 3D~fields~\cite{Pisanty:2019aa, Sugic:2020aa, Kessler:2003aa}. 

Our method is not limited to high harmonics. Its extension to low-order parametric processes such as chiral sum-frequency generation~\cite{Vogwell:2023aa} 
has potential 
for non-destructive enantiosensitive imaging in the UV region and for exploiting intrinsically low-order nonlinearities for enantiosensitive detection in the X-ray domain~\cite{Rouxel:2022aa, Rouxel:2022ab}.

\section*{Methods}

\subsection*{Spatial structure of vortex beams creating chiral topological light}

We use two Laguerre-Gaussian beams with  counter-rotating circular polarizations, propagating along the $z$-axis with frequencies $\omega$ and $2\omega$ and OAMs $\ell_\omega$ and $\ell_{2\omega}$. We set the radial indices to $p_\omega=p_{2\omega}=0$. The generalization to the case of non-zero radial index is straightforward. At the focal plane of the beams $z=0$, the Cartesian components of the fields in the transversal plane $(x,y)$ are
\begin{align}
\mathbf{E}^{\perp}_{\pm,r\omega}
&=
\mathcal{E}_{r\omega}e^{-\frac{\rho^2}{W^2_0}}\left(\frac{\sqrt{2}\rho}{W_0}\right)^{|\ell_{r\omega}|}e^{\text{i}\ell_{r\omega}\theta}e^{\text{i}\phi_{r\omega}}
\nonumber \\
&\qquad 
\times 
\frac{(\mathbf{e}_x-\text{i}\sigma_{r\omega}\mathbf{e}_y)}{\sqrt{2}}
,
\end{align}
where $\mathcal{E}_{r\omega}$
is the field strength, $W_0$ is the beam waist, $\phi_{r\omega}$ is the carrier-envelope phase (CEP), $\rho=\sqrt{x^2+y^2}$ and $\theta=\arctan(y/x)$ are the radial and azimuthal coordinates, and $\sigma_{r\omega}$ indicates right ($\sigma_{r\omega}=1$) or left ($\sigma_{r\omega}=-1$) circular polarization.
Near the focus this field develops a longitudinal component along the $z$-axis given by $E_z=-(\text{i}/k)\nabla_{\perp}\cdot\mathbf{E}_{\perp}$ in the first post-paraxial approximation~\cite{Bliokh:2015aa}:
\begin{align}
\mathbf{E}^z_{\pm,r\omega}
&=
-\frac{\text{i}\mathcal{E}_{r\omega}}{\sqrt{2}k_{r\omega}}e^{-\frac{\rho^2}{W^2_0}}\left(\frac{\sqrt{2}}{W_0}\right)^{|\ell_{r\omega}|}\rho^{|\ell_{r\omega}|-1}
\\\nonumber
& \qquad \times e^{\text{i}(l_{r\omega}+\sigma_{r\omega})\theta}\left(|\ell_{r\omega}|-\sigma_{r\omega}\ell_{r\omega}-\frac{2\rho^2}{W^2_0}\right)\mathbf{e}_z.
\end{align}
The total bichromatic electric field $\mathbf{E}(x,y)=\mathbf{E}_{\pm,\omega}+\mathbf{E}_{\pm,2\omega}$, combining the longitudinal and transverse field components for each color $\mathbf{E}_{\pm,r\omega}=\mathbf{E}^{\perp}_{\pm,r\omega}+\mathbf{E}^{z}_{\pm,r\omega}$ ($r=1,2$), is an example of a synthetic chiral light~\cite{Ayuso:2019aa}.

\subsection*{Chiral correlation function}

We report here the analytical expression for the chiral correlation function~\cite{Ayuso:2019aa} $h^{(5)}(-2\omega,-\omega,\omega,\omega,\omega)=\mathbf{E}^*(2\omega)\cdot\left[\mathbf{E}^*(\omega)\times\mathbf{E}(\omega)\right]\left(\mathbf{E}(\omega)\cdot\mathbf{E}(\omega)\right)$ for the general case of two OAM-carrying beams with frequencies $\omega$ and $2\omega$, SAMs $\sigma_{\omega}$ and $\sigma_{2\omega}$ and OAMs $\ell_\omega$ and $\ell_{2\omega}$. 
{\allowdisplaybreaks%
\begin{align}
h^{(5)}(\rho,&\theta)
=-\frac{\mathcal{E}_{2\omega}\mathcal{E}^4_\omega}{\sqrt{2}2k^2_\omega}e^{-5\frac{\rho^2}{W^2_0}}\left(\frac{\sqrt{2}}{W_0}\right)^{4|\ell_\omega|+|\ell_{2\omega}|}
\nonumber\\
&
\rho^{4|\ell_{\omega}|+|\ell_{2\omega}|-3}\left(|\ell_{\omega}|-\sigma_\omega\ell_\omega-\frac{2\rho^2}{W^2_0}\right)^2\nonumber\\
&
\left\{\frac{|\ell_\omega|-\sigma_\omega\ell_\omega-2\frac{\rho^2}{W^2_0}}{2k_\omega}\left[e^{i\sigma_\omega\theta}(\sigma_\omega-\sigma_{2\omega})-\nonumber\right.\right.\\
&
\left.e^{-i\sigma_\omega\theta}(\sigma_\omega+\sigma_{2\omega})\right]\nonumber\\
&
\left.+\frac{\sigma_{2\omega}}{k_{2\omega}}(|\ell_{2\omega}|-\sigma_{2\omega}\ell_{2\omega}-\frac{2\rho^2}{W^2_0})e^{-i\sigma_{2\omega}\theta}\right\}\nonumber\\
&
e^{i(2\phi_\omega-\phi_{2\omega})}e^{i(2\ell_\omega+2\sigma_\omega-\ell_{2\omega})\theta}
\end{align}
}%
It is easy to verify that both in the counter-rotating $\sigma_\omega=-\sigma_{2\omega}$ and co-rotating case $\sigma_\omega=\sigma_{2\omega}$ the azimuthal dependence of the chiral correlation function is given by $C\theta$, where $C=2(\ell_\omega+\sigma_\omega)-(\ell_{2\omega}+\sigma_{2\omega})$.

\subsection*{DFT-based SFA simulations in fenchone}

The method is adapted from Refs.~\cite{Ayuso:2019aa, Ayuso:2018aa, 
Smirnova:2014aa} to describe HHG in a chiral molecule subjected to a strong field. The macroscopic dipole moment in an ensemble of randomly oriented molecules arises form the coherent summation of the contributions from all possible molecular orientations
$$\mathbf{D}(N\omega)=\int d\Omega \int d\beta\,\mathbf{D}_{\Omega\beta}(N\omega) ,$$
where $\omega$ is the fundamental frequency, $N$ is the harmonic number and $\mathbf{D}_{\Omega\beta}$ is the harmonic dipole associated with a molecular orientation characterized by the three Euler angles, here denoted in terms of the solid angle $\Omega$ and the angle $\beta$. In the strong-field approximation~(SFA), the harmonic dipole for a given orientation~\cite{Ayuso:2019aa, Smirnova:2014aa} is given by 
\begin{eqnarray}\mathbf{D}_{\Omega\beta}(N\omega)=&&e^{\text{i}N\omega t'_r}a_{\mathrm{rec}}\,\mathbf{d}(\mathrm{U}_{\Omega\beta}\mathrm{Re}[\mathbf{k}(t'_r)])a_{\mathrm{prop}}\,\nonumber\\
&&e^{-\text{i}S(\mathbf{p}_s,t_i,t_r)}\,a_{\mathrm{ion}}\,\Psi_{D}(\mathrm{U}_{\Omega\beta}\mathrm{Re}[\mathbf{k}(t'_i)]),\end{eqnarray}
where $\mathbf{d}(\mathbf{k})$ is the recombination matrix element in the laboratory frame and $\mathbf{k}(t)=\mathbf{p}+\mathbf{A}(t)$. Here $\mathrm{U}_{\Omega\beta}$ is the rotation matrix that transforms the laboratory frame $(\mathbf{e}_1,\mathbf{e}_2,\mathbf{e}_3)$ to the molecular $(\mathbf{i}_1,\mathbf{i}_2,\mathbf{i}_3)$ frame, with elements $\mathrm{U}_{ij}=\langle\mathbf{e}_i|\mathbf{i}_j\rangle$ for a given orientation.Here $\Psi_D(\mathbf{k})=\langle\mathbf{k}|\Psi_D\rangle$ is the overlap between the Volkov state with kinetic momentum $\mathbf{k}$ and the Dyson orbital, where the latter is the overlap between the neutral $N$-electron wavefunction and the ionic $(N-1)$-electron wavefunction $|\Psi_D\rangle=\langle\Psi^{N-1}|\Psi^{N}\rangle$. The integral over the solid angle $d\Omega=d\alpha d\beta\sin(\beta)$ is performed using the Lebedev quadrature method~\cite{Lebedev:1975aa}, while the integral over the $\beta$ angle is done by trapezoid method. In order to find the rotation matrix, we first assume that the $x$-axis of the molecular frame points toward a given Lebedev point, and then rotate by an angle $\beta$ around the $x$-axis. For all simulations we use a 17th-order Lebedev quadrature (for a total of 110 points) and 40 $\beta$ angles evenly distributed on the $[0,2\pi]$ interval.

In the expression for the harmonic dipole, $\mathbf{p}$, $t_i=t'_i+\text{i}t''_i$, $t_r=t'_r+\text{i}t''_r$ are the complex momenta and times of ionization and recombination resulting from the application of the saddle-point method~\cite{Smirnova:2014aa}. $S(\mathbf{p},t_i,t_r)=\frac12 \int_{t_i}^{t_r}dt'\,\left[\mathbf{p}+\mathbf{A}(t')\right]^2+I_p(t_r-t_i)$ is the action from the (complex) times of ionization and recombination. The terms associated with the saddle-point method on $(t_i,t_r,\mathbf{p})$ are given by
\begin{align}
a(\mathbf{p},t_i,t_r)&=a_\mathrm{ion}a_\mathrm{prop}a_\mathrm{rec}
\nonumber \\
a_\mathrm{ion}&=\sqrt{\frac{2\pi}{\partial^2_{t_i}S}}
\nonumber \\
a_\mathrm{rec}&=\sqrt{\frac{2\pi}{\partial^2_{t_r}S}}
\nonumber \\
a_\mathrm{prop}&=\left(\frac{2\pi}{\text{i}(t_r-t_i)}\right)^{3/2}
\nonumber 
\end{align}
where the second derivatives of the action are given explicitly by
\begin{align}
\partial^2_{t_i}S=-\mathbf{E}(t_i)\cdot\mathbf{k}(t_i),
\nonumber \\
\partial^2_{t_i}S=\mathbf{E}(t_r)\cdot\mathbf{k}(t_r),
\nonumber 
\end{align}
where $\mathbf{E}(t)$ is the electric field
and all expressions for the prefactor are calculated at the complex times.

The transition matrix elements of the right- and left-handed molecules are related by 
\begin{equation}
\mathbf{D}_R(\mathbf{k})=-\mathbf{D}_L(-\mathbf{k}),
\end{equation}
while for the overlap between the Dyson orbital and the Volkov wavefunction we have that
\begin{equation}\Psi^R_D(\mathbf{k})=\Psi^L_D(-\mathbf{k}).\end{equation}
The matrix elements and the Dyson orbitals for fenchone are calculated using DFT methods described in~\cite{Ayuso:2021aa, TOFFOLI200225}.

\subsection*{Multiphoton picture}

The multiphoton picture of enantiosensitive HHG driven by chiral topological light can be understood by analyzing the contributing chiral and achiral multiphoton pathways. To do so, we classify the multiphoton pathways by indicating with a subscript the SAM of the photon, so that e.g.\ $(N)\omega_+$ indicate the absorption of $N$ $\omega$ photons with SAM $m=1$ and $(-1)\omega_z$ indicates the emission of one $\omega$ photon with SAM $m=0$.

In the specific case of bicircular counter-rotating fields, if the field has no longitudinal component along its direction of propagation (i.e.\ if we consider an achiral field in the dipole approximation), conservation of SAM results in a harmonic spectrum with doublets at $3N+1$ and $3N+2$ harmonic frequencies, where the $3N+1$ harmonics ($3N+2$) co-rotate with the $\omega$ ($2\omega$) field~\cite{Fleischer:2014aa, Hickstein:2015aa}. $3N$ harmonic orders are forbidden in achiral media, since their generation requires absorption of an equal number of photons from both drivers. In chiral media, the $3N$ harmonic orders can instead be generated due to the broken parity of the medium, but are polarized along the direction of propagation of the fields (the $z$-axis in our case), and thus are not detectable in the far-field. We label this pathway as
\begin{equation}C_z=\left[(N)\omega_+,(N)2\omega_-\right].\end{equation}
Focusing on the specific case of $3N$ harmonic orders, if the field is chiral (i.e.\ if it posses a longitudinal component along the propagation direction) in the case of achiral media the following multiphoton pathways can now lead to symmetry-allowed HHG:
\begin{align}
\mathrm{AC}_+&=\left[(N-2)\cdot\omega_+, (2)\omega_z, (N-1)\cdot2\omega_-\right]\\
\mathrm{AC}_-&=\left[(N-1)\cdot\omega_+,(-1)\omega_z, (N)\cdot2\omega_{-}, (1)2\omega_z\right]
\end{align}
corresponding respectively to the emission of a photon with SAM $m=1$ and $m=-1$. We label these pathways as achiral pathways (i.e.\ $\mathrm{AC}_{m}$, with $m$ the SAM of the harmonic photon), since they occur already in achiral media driven by a chiral field as they require the absorption and emission of an odd number of photons. If the medium is chiral, two new pathways including absorption of an equal number of $\omega$ and $2\omega$ photons open, i.e.\
\begin{align}\mathrm{C}_+&=\left[(N)\cdot\omega_+, (N-1)\cdot2\omega_-, (1)2\omega_z\right]\\
\mathrm{C}_-&=\left[(N-1)\cdot\omega_+,(1)\omega_z, (N)\cdot2\omega_{-}\right]\end{align}
corresponding again respectively to the emission of a photon with SAM $m=1$ and $m=-1$. We label these pathways as chiral pathways ($\mathrm{C}_m$) since they can occur only in chiral media. Finding the corresponding OAM of all pathways indicated above is straightforward, once we remember that the longitudinal components of the fields carry OAMs of $\ell_{\omega_z}=\ell_{\omega_+}+\sigma_{\omega}$ and $\ell_{2\omega_z}=\ell_{2\omega_-}+\sigma_{2\omega}$. Obviously, other chiral and achiral pathways including the absorption of a larger number of $z$-polarized photons from either drivers are also in principle accessible: yet, since the longitudinal component is relatively weak, we restrict ourselves here to the photon pathways that include the absorption or emission of the fewest number of $z$-polarized photons. Fig.~1a of the SI) shows schematically the multiphoton pathways $\mathrm{C}_z$, $\mathrm{AC}_{m}$ and $\mathrm{C}_m$ for the case of a $3N$ harmonic order.

The results from the SFA simulations confirm the considerations above; in Fig.~1b  of the SI we show the near-field OAM distributions for H18 in R-fenchone driven by a field with $\ell_\omega=-\ell_{2\omega}=1$ and $\sigma_{\omega}=-\sigma_{2\omega}=1$. For comparison, we also report the OAM content for an artificial atom with ionization potential equal to fenchone driven by the same chiral field and the OAM content in fenchone for an achiral field with same OAM of the driving beams, obtained by manually setting the longitudinal component of the field to zero.\\
When the field is achiral, H18 in an atom is absent, while in the case of fenchone we observe a $\ell=0$ component polarized along the $z$-axis: this corresponds to the pathway $\mathrm{C}_z$ denoted above. When the field is chiral, circularly polarized components with $\ell=\pm5$ are observed for both the atom and the molecule: these are the achiral pathways $\mathrm{AC}_{+}$ and $\mathrm{AC}_{-}$ denoted above. Finally, the chiral pathways $\mathrm{C}_{+}$ and $\mathrm{C}_{-}$ correspond to the OAMs $\ell=\pm1$ and are only seen in a chiral molecule, since they require the absorption of an even number of photons.
Note that in the far-field only the SAM $m=\pm1$ components are going to be observed, since $m=0$ polarization (corresponding to the $\mathrm{C}_z$ pathway in black in Fig.~1b of the SI) will propagate in a direction orthogonal with respect to the propagation axis of the beams.

The different OAM content of an atom and chiral molecule driven by a chiral bicircular field is directly reflect in the far-field profile of H18, shown in Fig.~1c of the SI). In an atom (left figure of Fig.~1c of the SI), where for a given SAM there is only one contributing OAM, the far-field profile of H18 is a ring where the intensity is mostly constant, while in fenchone we observe an azimuthal interference pattern with periodicity determined by the topological charge $C$, corresponding in modulus to the net difference between the OAMs of chiral and achiral pathways. The enantiosensitive rotation of the spatial profile can be understood from the perspective of the multiphoton pathways by accounting a shift by $\pi$ of the phase of the chiral pathways $\mathrm{C}_\pm$ when changing the molecular enantiomer.
The enantiosensitive rotation of the spatial profile of the high harmonics in the far-field allows one also to use HHG driven by chiral vortices as a highly-sensitive method to infer the enantiomeric excess in a mixture of right- and left-handed molecular enantiomers.

Next order pathways can be identified using the same approach. In the case of achiral channels the next order pathway includes the absorption of two more longitudinal photons (see Fig.~2 of the SI) and is respectively two order of magnitude smaller. The next order chiral pathway is four order of magnitude smaller, corresponding to the absorption of four more longitudinal photons, and so on.

As mentioned in the main text in the case of an elliptically polarized $\omega$ field two new achiral pathways dominate the response, whose photon diagrams we report in Fig.~3 of the SI. For a $3N$ harmonic order both new achiral pathways contribute to the final SAM of $m=-1$ and are in particular
\begin{align}
\mathrm{AC}^{\epsilon}_{1}&=\left[(N-2)\cdot\omega_+,(2)\omega_-,(N-1)\cdot2\omega_-\right]\\
\mathrm{AC}^\epsilon_2&=\left[(N-1)\cdot\omega_+,(-1)\omega_-,(N+1)\cdot2\omega_-\right],
\end{align}
where $\omega_-$ refers now to the counter-rotating component of the elliptically polarized field at $\omega$ frequency. Since each elliptically polarized photon carries a phase delay dependence of $\exp(\text{i}\delta)$, the interference between these two achiral pathways oscillates with respect to the phase delay as $3\delta$. This explains why choosing the $\tilde{\delta}=1$ component of the harmonic profile after Fourier analysis allows one to recover the enantiosensitive rotation of the spatial profile.

\subsection*{Noise (intensity fluctuations) simulations}

In order to include the effect of noise on HHG driven by chiral vortex light, we take the following approach.
For a given electric field strength $E_0$ (which we assume to be the same for both fields) the Laguerre-Gaussian beam is given in the near-field by $\mathbf{E}(\mathbf{r})=E_0\mathbf{LG}_{l,p}(\mathbf{r})$, where $\mathbf{LG}_{l,p}=LG_{l,p}(\mathbf{r})\mathbf{e}_L(\mathbf{r})$. Here $LG_{l,p}$ is a Laguerre-Gaussian mode and $\mathbf{e}_L$ is the polarization vector of the field. The corresponding laser intensity is $I_0=|E_0|^2$. We then pick a value for the laser intensity from a normal distribution of noise centered at $I_0$ with width $\gamma$. We call this electric field intensity $I_1$. Then, for each point $\mathbf{r}$ in the focus, we introduce intensity fluctuations such that at a given position the electric field strength is given by
\begin{equation}\mathcal{I}(\mathbf{r})=I_1\,LG_{l,p}(\mathbf{r})(1+\delta_I(\mathbf{r})),\end{equation}
where $\delta_I(\mathbf{r})=C\lambda(\mathbf{r})$. $\lambda(\mathbf{r})$ is chosen from a Gaussian distribution centered at zero with width 1 and $C=0.1$ is a constant. There is therefore 68.2\% probability that the fluctuation is below 0.1\% of the signal at the given point. We produce 16 electric fields using this approach, choosing a central intensity of $I_0=5\cdot10^{14}$ W/cm$^2$ with width $\gamma=3.51\cdot10^{13}$ W/cm$^2$, and calculate the resulting far-field picture for left- and right-handed fenchone. The average intensity fluctuations are on the order of 2\%, on par with standard experimental parameters~\cite{Astrella}. We then scan the enantiomeric excesses $ee$ between $-100\%$ and $100\%$ in 1001 steps. For each step, we pick a random index $i$ between 1 and 16, selecting one of the far-field profiles for R- and L-fenchone $\mathbf{d}^{R/L}_i$. The resulting far-field image at a given enantiomeric excess $ee=(N_R-N_L)/(N_R+N_L)$ for normalized concentrations $N_R+N_L=1$ is given by $\mathbf{d}^{ee}=N_R\mathbf{d}^{R}_i+N_L\mathbf{d}^{L}_i$ and the phase of the $\ell=6$ Fourier component of the outer ring $kW_0>10$ is then calculated. We then repeat the procedure 16 times and for each enantiomeric excess calculate the mean phase as $\bar{\phi}=\sum_{i=1}^{16}\phi_i/16$. The result is the red solid line shown in Fig.~\ref{Fig3}e.

\section*{Data availability}

The data used to generate the figures resulting from the numerical simulations in this paper has been publicly archived in the Zenodo repository at \href{https://doi.org/10.5281/zenodo.11501346}{https://doi.org/10.5281/zenodo.11501346}.

\section*{Code availability}

The code used to generate the data is available from the authors on reasonable request.

\section*{Acknowledgements}

The authors would like to acknowledge helpful discussions with A.~Ord\'{o}\~{n}ez and O.~Kornilov.
O.S.\ and M.I.\ acknowledge the hospitality of the Technion~-- Israel Institute of
Technology, especially during the week commencing October 8, 2023. This
project has received funding from the EU Horizon 2020 programme (grant 
agreement No 899794) and European Union (ERC, ULISSES, 101054696).
D.A., M.K.\ and E.P.\ acknowledge Royal Society funding under 
URF\textbackslash{}R1\textbackslash{}201333, URF\textbackslash{}R1\textbackslash{}231460, and URF\textbackslash{}R1\textbackslash{}211390.

\section*{Author contributions}

N.M.\ and O.S.\ developed the concept of chiral topological light and its application to enantiosensitive spectroscopy. N.M.\ computed and analyzed all microscopic and macroscopic HHG response using the DFT-based SFA simulations developed by D.A.\ and P.D. N.M.\ and O.S.\ wrote the initial version of the manuscript. All authors contributed to writing the manuscript.

\balance


\bibliographystyle{arthur} 
\bibliography{references}{}

\hfill 

\end{document}